\documentclass[12pt]{article}
\usepackage{amssymb,bbold}
\usepackage{graphicx}
\usepackage[hang,small,bf]{caption}
\usepackage{hyperref}

\newcommand{\be}[3]{\begin{equation}  \label{#1#2#3}}
\newcommand{\ee}{\end{equation}}
\newcommand{\ba}{\begin{array}}
\newcommand{\ea}{\end{array}}
\newcommand{\bea}[3]{\begin{eqnarray}  \label{#1#2#3}}
\newcommand{\eea}{\end{eqnarray}}

\let\LARGE=\Large
\let\Large=\large
\let\large=\normalsize

\def\1{\mathbb 1}
\def\F{{\cal F}^{(1)}}
\def\FF{{\cal F}^{(7)}}
\def\FFF{{\cal F}^{(27)}}

\newcommand{\haken}{\mathbin{\hbox to 8pt{%
                 \vrule height0.4pt width7pt depth0pt
                 \kern-.4pt
                 \vrule height4pt width0.4pt depth0pt\hss}}}

\newcommand{\resetcounter}{\setcounter{equation}{0}}  

\begin{document}

\baselineskip=20pt
\parskip=6pt


\thispagestyle{empty}

\begin{flushright}
\hfill{AEI-2003-093}\\
\hfill{HU-EP-03/77} \\
\hfill{hep-th/0311119}

\end{flushright}

\vspace{10pt}

\begin{center}{ \LARGE{\bf
Fluxes in M-theory on 7-manifolds: \\[4mm]
$G$-structures and Superpotential
}}

\vspace{35pt}

{\bf Klaus Behrndt}$^a$ \quad and \quad
{\bf Claus Jeschek}$^b$

\vspace{15pt}

$^a$ {\it  Max-Planck-Institut f\"ur Gravitationsphysik,
Albert Einstein Institut\\
Am M\"uhlenberg 1,  14476 Golm,
 Germany}\\[1mm]
{E-mail: behrndt@aei.mpg.de}

\vspace{8pt}

$^b$ {\it  Humboldt Universit\"at zu Berlin,
Institut f\"ur Physik,\\
Invalidenstrasse 110, 10115 Berlin,
 Germany}\\[1mm]
{E-mail: jeschek@physik.hu-berlin.de}

\vspace{40pt}

{ABSTRACT}

\end{center}

\noindent
We consider compactifications of M-theory on 7-manifolds in the
presence of 4-form fluxes, which leave at least four supercharges
unbroken.  We focus especially on the case, where the 7-manifold
supports two spinors which are SU(3) singlets and the fluxes appear as
specific SU(3) structures. We derive the constraints on the fluxes
imposed by supersymmetry and calculate the resulting 4-dimensional
superpotential.

\vfill

\newpage


\section{Introduction}


Compactifications of string or M-theory in the presence of fluxes are
expected to improve a number of problems appearing in generic
4-dimensional field theories obtained by Kaluza-Klein reductions.  In
order to get contact to the standard model, one has to solve e.g.\ the
moduli problem, the problem of missing chiral fermions and the gauge
hierarchy problem. We refer to non-trivial fluxes as 10- or
11-dimensional gauge field strengths that are non-zero in the
vacuum\footnote{Note, we are using the flux/brane notation in a rather
loose sense and for other applications one might relate fluxes only to
closed forms whereas fluxes coming from (magnetic) branes are related
to source terms in the Bianchi identities.}  and hence give a
potential upon dimensional reduction, which lift at least part of the
moduli space. So far it is unclear whether all moduli can be lifted in
this way -- especially not if one wants to preserve some
supersymmetries.  Concerning chiral fermions, the situation was
hopeless for long time and only recently it was realized that chiral
fermions can appear on the world volume of D-branes; e.g.\ on
D3-branes at specific singularities in the internal space or on
intersecting D6-branes. Since D-branes are sources for fluxes,
(intersecting) D-brane configurations represent one example of flux
compactifications. One should of course distinguish between branes
which produce the background geometry (e.g.\ $AdS_5 \times S_5$) and
localized probe branes which are assumed to produce no back reaction
on the geometry. Another feature of flux compactifications is the
non-trivial warping yielding a suppression of the bulk (gravitational)
scale in comparison to the world volume gauge theory scale and has
therefore been discussed with respect to the gauge hierarchy problem.

By now one can find a long list of literature about this subject.  A
starting point was the work by Candelas and Raine \cite{480} for an
un-warped metric which was generalized later in \cite{490} (for an
earlier work on warp compactification see \cite{500}) and the first
examples, which preserve N=1 supersymmetry appeared in \cite{520,
230}. The subject was revived around 10 years later by the work of
Polchinski and Strominger \cite{300}, where flux compactifications in
type II string theory was considered. Since then many aspects have
been discussed \cite{530}--\cite{570};
we will come back to some highlights of these papers.

Recall, a somewhat trivial example of flux compactifications are
(intersecting) brane configurations where one assumes that the
external space is part of the (common) worldvolume directions; as
e.g.\ the intersecting M5-brane configurations \cite{100, 460,
290,200}.  As the branes have to follow certain intersection rules
imposed by supersymmetry also the fluxes have to satisfy certain
constraints, i.e.\ some components have to be zero and others will fix
the warp factor. At the same time, fluxes induces a non-trivial back
reaction onto the geometry and this back reaction can be made explicit
by re-writing the flux components as specific con-torsion components,
see e.g.\ \cite{250, 330, 240, 170, 260, 140}.  The resulting spaces
are in general non-K\"ahlerian and their moduli spaces are
unfortunately only purely understood.  On the other hand, the fixing
of moduli can be addressed by deriving the corresponding
superpotential as function of the fluxes in a way discussed in
\cite{160,280,410}. But also here one observes difficulties.
Non-trivial Bianchi identities exclude an expansion of the fluxes in
terms of harmonic forms, which in turn makes it difficult to decide to
what extend the moduli space is lifted, see \cite{330}.

In this paper we will consider M-theory compactifications in the
presence of 4-form fluxes, which keep the external 4-d space time
maximal symmetric, i.e.\ either flat or anti deSitter (AdS), where in
the latter case the superpotential remains non-zero in the vacuum. In
many papers the superpotential is derived directly by Kaluza-Klein
reduction, whereas we introduce it as a mass term for the 4-d
gravitino, i.e.\ a (non-vanishing) covariant derivative of the 4-d
Killing spinors. In general one has not only a single internal and
external spinor, which gives a classification of the solutions by the
number of spinors. One can have up to eight spinors, but this case is
technically very involved and severely constrains the internal space,
see \cite{230}. On the other hand, one can consider first the single
spinor case, followed by the two-spinor case, the three-spinor case
and so on. These spinors are singlets under the structure group of the
corresponding manifold and fluxes will introduce the corresponding
$G$-structures (or torsion components). Thus, this is a classification
of the vacua with respect to the $G$-structure group, which is a
subgroup of SO(7). The single spinor case which should allow for
general $G_2$-structures is rather trivial, see also \cite{130}.  The
first non-trivial case is given by two internal spinors which give
rise to SU(3)-structure and which is the main subject of present
paper; see also \cite{110} for a discussion for this case without
superpotential.  The reduction of the structure group is a consequence
of the appearance of a vector field built as a fermionic
bi-linear. But this is not the most general case. Since on any
7-manifold, that is spin, exist up to three independent vector fields
\cite{380}, one can always define SU(2)-structures. We shall add here
a warning. Any vacuum which preserves four supercharges, (i.e. $N$=1,
D=4), should finally be described by one 4-d spinor and one 7-d
spinor, but the 11-d spinor is in general not a direct product of
these two spinors. In fact, the 7-d spinor will exhibit a
$\gamma^5$-dependent rotation while moving on the 7-manifold, see also
\cite{382}. In our counting we refer to the maximal number of
independent spinors to the number of singlet spinors under a given
structure group and the resulting projector constraints give us a
powerful tool to solve the Killing spinor equations. At the end of the
day, the different 4-d spinors might be related to each other so that
we still describe $N$=1 vacua.

We have organized the paper as follows. In the next section we will
make the ansatz for the metric and the 4-form field strength and
separate the gravitino variation into an internal and external part.
In order to solve these equations, we have to decompose the 11-d
spinor into an external and internal part. This is done in Section 3.
Using a global vector field, which defines a foliation of the
7-manifold $X_7$ by a 6-manifold $X_6$, we define SU(3) singlet
spinors on $X_7$. The corresponding projector equations are given in
detail and can be used to re-write the flux contribution as
con-torsion components, which is discussed in Section 4, where we also
introduce the superpotential as mass term for the 4-d gravitino.
After having introduced all relevant relations, we will derive in
Section 5 the supersymmetry constraints on the fluxes and the warp
factor and we distinguish especially between the single and 2-spinor
case.  These sections are rather technical and the reader more
interested in the results might go directly to Section 6, where we
give a discussion of our results after projecting them on $X_6$ (i.e.\
after making a SU(3) decomposition). The M-theory 4-form $F$
decomposes on $X_6$ into a 3-form $H$ and a 4-form $G$.  The
holomorphic (3,0)-part of $H$ defines the 4-d superpotential, the
non-primitive (2,1)-part fixes the warp factor whereas the primitive
part is un-constrained.  The 4-form $G$ has to be of (2,2)-type on
$X_6$.  Finally in Section 7 we comment on special cases: (i) if the
vector field is closed; (ii) if it is Killing, (iii) the vector is
Killing, but with non-constant norm and finally we comment in (iv) on
the MQCD-brane world model as discussed by Witten \cite{100}.

{\bf Note added.} While preparing this paper, we came aware of the
work done by Dall'Agata and Prezas \cite{470}, which have some overlap
with our results, but discuss in more detail the reduction to type IIA
string theory.


\section{Warp compactification in the presence of fluxes}
\resetcounter


The compactifications of M-theory in the presence of 4-form fluxes
imply in the generic situation not only a non-trivial warping, but
yield a 4-d space time that is not anymore flat. This is a consequence
of the fact, that for generic supersymmetric compactifications, the
fluxes generate a superpotential, that is non-zero in the vacuum and
the resulting (negative) cosmological constant implies a 4-d anti
deSitter vacuum.  Note, we are not interested in the situation where
the 4-d superpotential exhibits a run-away behavior, i.e.\ has no
fixed points. We consider therefore as ansatz for the metric and the
4-form field strength
\be012
\ba{rcl}
ds^2 &=& e^{2 A} \Big[ \, g^{(4)}_{\mu\nu} dx^\mu dx^\nu
        + \, h_{ab} dy^a dy^b \Big] \ , \\
F &=& {m \over 4!} \,  
\epsilon_{\mu\nu\rho\lambda} dx^\mu \wedge dx^\nu \wedge
    dx^\rho \wedge dx^\lambda +
{1 \over 4!} F_{abcd} \, dy^a \wedge dy^b \wedge dy^c \wedge dy^d
\ea
\ee
where $A=A(y)$ is a function of the coordinates of the 7-manifold with
the metric $h_{ab}$, $m$ is the Freud-Rubin parameter and the 4-d
metric $g^{(4)}_{\mu\nu}$ is either flat or anti deSitter.

Unbroken supersymmetry requires the existence of (at least) one
Killing spinor $\eta$ yielding a vanishing gravitino variation of
11-dimensional supergravity
\be716
\ba{rcl}
0 = \delta \Psi_M &=& 
        \Big[ \partial_M + {1 \over 4} \hat \omega^{RS}_M \Gamma_{RS}
        + {1 \over 144} \Big(\Gamma_M^{\ NPQR} - 8 \, \delta_M^N\, 
        \Gamma^{PQR} \Big) \, F_{NPQR} \Big] \eta \\
&=& 
        \Big[ \partial_M + {1 \over 4} \hat \omega^{RS}_M \Gamma_{RS}
        + {1 \over 144} \Big(\Gamma_M \hat F - 12 \, \hat F_M \Big) \Big] \eta 
\ .
\ea
\ee
In the second line we used the formula
\be371
\Gamma_M \Gamma^{N_1 \cdots N_n} = \Gamma_M^{\ N_1 \cdots N_n} +
n \, \delta_M^{\ [N_1} \Gamma^{N_2 \cdots N_n]}
\ee
and introduced the abbreviation
\be201
\hat F \equiv F_{MNPQ} \Gamma^{MNPQ} \quad, \qquad
\hat F_M \equiv F_{MNPQ} \Gamma^{NPQ}  \ .
\ee
Using the convention $ \{ \Gamma^A , \Gamma^B \} = 2 \eta^{AB}$ with
$\eta = {\rm diag}(-,+,+ \ldots +)$, we decompose the $\Gamma$-matrices
as usual
\be391
\Gamma^\mu = \hat \gamma^\mu \otimes \1 \qquad , \qquad
\Gamma^{a+3} = \hat \gamma^5 \otimes \gamma^a
\ee
with $\mu = 0,1,2,3$, $a = 1,2, \ldots 7$ and
\be791
\hat \gamma^5 = i \hat \gamma^0 \hat \gamma^1 \hat \gamma^2 \hat 
\gamma^3 \ , \quad  \gamma^1  \gamma^2  \gamma^3 \gamma^4 \gamma^5 
         \gamma^6  \gamma^{7} = - i 
\ee
which implies
\be161
i \hat \gamma^5 \hat \gamma^\mu = {1 \over 3!} \epsilon^{\mu\nu\rho\lambda}
\hat \gamma_{\nu\rho\lambda} \quad ,\qquad
{i \over 3!} \epsilon^{abcdmnp} \gamma_{mnp} =
\gamma^{abcd} \equiv \gamma^{[a} \gamma^b \gamma^c \gamma^{d]} \ .
\ee
The spinors in 11-d supergravity are in the Majorana representation
and hence, all 4-d $\hat \gamma^\mu$-matrices are real and $\hat
\gamma^5$ as well as the 7-d $\gamma^a$-matrices are purely imaginary
and antisymmetric.

With this notation, we can now split the gravitino variation into an
internal and external part. First, for the field strength we find
\be018
\ba{l}
\hat F = - i \, m \, \hat \gamma^5 \otimes {\mathbb 1} + 
{\mathbb 1} \otimes F\ , 
\\
\hat F_\mu = {1 \over 4} \, i \, m  \hat \gamma^5 \hat \gamma_\mu \otimes 
   {\mathbb 1} \quad , \qquad \hat F_a = \hat \gamma^5 \otimes F_a
\ea
\ee
where $F$ and $F_a$ are defined as in (\ref{201}), but using the 7-d
$\gamma^a$-matrices instead of the 11-d matrices.  In order to deal
with the warp factor, we use
\be625
ds^2 = e^{2A} \widetilde{ds}^2 \quad \rightarrow \quad
D_M = \tilde D_M + {1 \over 2}    \Gamma_M^{\ N} \partial_N A
\ee
and find for the external components of the gravitino variation
\be112
0=\Big[ \nabla_\mu \otimes \1 +  \hat \gamma_\mu \hat \gamma^5
  \otimes \Big( {1 \over 2} \, \partial A + {i \, m \over 36} \Big)
+ {1 \over 144} e^{-3A} \, \hat \gamma_\mu \otimes F
  \Big] \eta  
\ee
where $\partial A \equiv \gamma^a \partial_a A$ and $\nabla_\mu$ is the
4-d covariant derivative in the metric $g^{(4)}_{\mu\nu}$.
In the same way, we get for the internal gravitino variation
\be836
0 = \Big[ \1 \otimes \Big( \nabla_a^{(h)} + {1 \over 2}
  \gamma_a^{\ b} \, \partial_b A + {i\, m  \over 144} \, \gamma_a \Big) + 
        {1 \over 144} e^{-3A}\,  \hat \gamma^5 \otimes
  \Big( \gamma_a F  -12 F_a \Big) \Big] \eta \ .
\ee
Note, from this equation we can eliminate the term $\sim \gamma_a F \eta$
by multiplying eq.\ (\ref{112}) with (${1 \over 4} \hat \gamma^5 \hat
\gamma^\mu\otimes \gamma^a$) and subtracting both expression.
As result we obtain
\be711
0= \Big[ \1 \otimes
\Big( \nabla_a^{(h)} -  {1 \over 2} \partial_a A  
+ { i m \over 48} \gamma_a \Big)  - 
{1\over 4} \hat \gamma^5  \hat \gamma^\mu \nabla_\mu \otimes \gamma^a -
{1 \over 12}\, e^{-3A}\, \hat \gamma^5 \otimes F_a \Big) \Big] \eta \ .
\ee
Before we can continue in solving this equation, we shall decompose
the spinor and introduce the superpotential.


\section{Decomposition of the Killing spinor}
\resetcounter


The 11-d Majorana spinor can be expanded in all independent spinors
\[
\eta = \sum_{i=1}^N (\epsilon^i \otimes \theta^i + cc) \ .
\]
where $\epsilon^i$ and $\theta^i$ denote the 4- and 7-d spinors,
respectively.  If there are no fluxes, all of these spinors are
covariantly constant and $N$ gives the number of extended
supersymmetries in 4 dimensions, where $N=8$ is the maximal possible.
In presence of fluxes, the spinors are not independent anymore and
hence $N$ does not refer to the number of unbroken supersymmetries,
but nevertheless, this number gives a classification of the
supersymmetric vacua. Basically, this is a classification with respect
to the subgroup of SO(7) under which the spinors are invariant and the
embedding of this subgroup is parameterized by globally well-defined
vector fields.  If there is e.g., only a single spinor on the
7-manifold, it can be chosen as a real $G_2$ singlet spinor; if there
are two spinors, one can combine them into a complex $SU(3)$ singlet;
three spinors can be written as $Sp(2)\simeq SO(5)$ singlets and four
spinors as $SU(2)$ singlets.  Of course, all eight spinors cannot be a
singlet of a non-trivial subgroup of $SO(7)$.  Apart from the latter
case, the internal spinors satisfy certain projectors, which
annihilate all components that would transform under the corresponding
subgroup and, at the same time, these projectors can be used to derive
simple differential equations for the spinors (as we will discuss in
more detail in the next section). For vanishing superpotential a
discussion of the single spinor case is given in \cite{130} whereas
the two spinor case has been discussed before in \cite{110}.

These spinors should be globally well-defined on the 7-manifold, and
hence, they can be used to build differential forms as bi-linears
\[
\theta^i \gamma_{a_1 \cdots a_n} \theta^j \ .
\]
The 7-dimensional $\gamma$-matrices are in the Majorana representation
and satisfy the relation: $(\gamma_{a_1 \cdots a_n})^T =(-)^{n^2 +n
\over 2}\gamma_{a_1 \cdots a_n}$, implying that the differential form
is antisymmetric in $[i,j]$ if $n=1,2,5,6$ and otherwise symmetric [we
assumed here of course that $\theta^i$ are commuting spinors and the
external spinors are anti-commuting].  This gives the well-known
statement that a single spinor does not give rise to a vector or
2-form, but only a 3-form and its dual 4-form [the 0- and 7-form exist
trivially on any spin manifold]. If we have two 7-spinors
$\theta^{\{1/2\}}$, we can build one vector $v$ and one 2-form (and of
course its dual 5- and 6-form). Since the spinor is globally
well-defined also the vector field is well defined on $X_7$ and it can
be used to obtain a foliation of the 7-d space by a 6-manifold $X_6$,
see figure for a simple example. Similarly, if there are three
7-spinors we can build three vector fields as well as three 2-forms and
having four spinors the counting yields six vectors combined with six
2-forms. In addition to these vector fields and 2-forms, one obtains
further 3-forms the symmetrized combination of the fermionic
bi-linears. We have however to keep in mind, that all these forms are
not independent, since Fierz re-arrangements yield relations between
the different forms, see \cite{250, 240,140} for more details.

\begin{figure} 
\begin{center}
\includegraphics[angle=0,width=80mm]{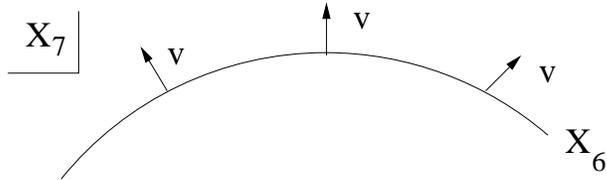}
\end{center}
\caption{We have shown here a simple case, where the vector $v$ built
out of two 7-d spinors gives a foliation of the 7-manifold $X_7$ by a
6-manifold $X_6$. This foliation is not unique since on any 7-manifold
that is spin exists at least three globally well-defined vectors.}
\label{fig1}
\end{figure}

In the simplest case with a single spinor, this $G_2$ singlet can be
written as
\be653
\theta  = e^{Z} \theta_0
\ee
where $\theta_0^T$ is a normalized real spinor obeying $\theta_0^T
\theta_0 = 1$, e.g.\ $\theta_0^T = (1,0, \ldots , 0)$.  In this case,
the 11-d spinor is written as
\be251
\eta = \epsilon \otimes \theta
\ee
and since the 11- and 7-d spinor are Majorana also the 4-d spinor
$\epsilon$ has to be Majorana.  As we will see below, the differential
equation for this $G_2$ singlet spinor becomes $\nabla_a \theta \sim
\gamma_a \theta$, which implies that $X_7$ is a space of weak $G_2$
holonomy. This in turn implies, that the 8-d space built as a cone
over this 7-manifold has $Spin(7)$ holonomy. As we will see below, 
the internal fluxes in this case have to be trivial.

Next, allowing for two spinors on $X_7$, one can build a vector field
$v$, which can be used to combine these two spinor into one complex
spinor defined as
\be726
\theta = {1 \over \sqrt{2}} \, e^Z \, ( \1 + v_a \gamma^a ) \theta_0
\quad , \qquad v_a v^a = 1
\ee
where the constant spinor $\theta_0$ is again the $G_2$ singlet and
$Z$ is a complex function.  Both spinors, $\theta$ and its complex
conjugate $\theta^\star$, are chiral spinors on $X_6$ (see figure) and
moreover, both spinors are $SU(3)$ singlets, which will become clear
from the properties that we will discuss below. In this case, the 11-d
Majorana spinor is decomposed as
\be010
\eta = \epsilon \otimes \theta + \epsilon^\star \otimes \theta^\star \ .
\ee
Of course, now $\epsilon$ is not a 4-d Majorana spinor, which would
bring us back to the single spinor case discussed before ($\theta +
\theta^\star \sim \theta_0$). Instead $\epsilon$ and $\epsilon^\star$
are two 4-d Weyl spinors of opposite chirality.  Now, the differential
equation for this 7-d spinor of the form $\nabla_a \theta \sim
\gamma_a \theta$ identifies $X_7$ as a Sasaki-Einstein space, 
which implies that the cone over it is a 8-d space of SU(4)-holonomy.

The other cases become now technically more and more involved.  If
there are three different spinors on $X_7$, we can build three vectors
as well as three 2-forms and the three spinors are now $Sp(2)$
singlets. One might think that the existence of three well-defined
vector fields on $X_7$ would give restrictions on the manifold, but
this is not the case. On any 7-manifold that is spin, exists at least
three no-where vanishing vector fields, \cite{380}.  The manifold can
again be identified from the equation satisfied by the spinors. The
equation $\nabla_a \theta^i \sim \gamma_a \theta^i$ implies e.g., that
$X_7$ a 3-Sasaki-space (i.e.\ the cone yields an 8-d Hyperk\"ahler
space with $Sp(2)$ holonomy), with the Aloff-Walach space $N^{1,1}$ as
the only regular examples \cite{360} (apart from $S^7$); non-regular
examples are in \cite{390}.

The four spinors case yield up to six vector fields (recall, due to
Fierz re-arrangements they are not necessarily independent), which
would however severely constrain the manifold (e.g.\ that $X_7$
factorizes into $R_3 \otimes X_4$).

Let us now return to the spinor $\theta_0$, which is fixed by the
requirement to be a $G_2$ singlet.  Note, the Lie algebra
$\mathfrak{so}(7)$ is isomorphic to $\Lambda^2$ and  a reduction of the
structure group on a general $X_7$ from $SO(7)$ to the subgroup $G_2$
implies the following splitting:
\be980
\mathfrak{so}(7) \, = \, \mathfrak{g}_2 \, \oplus \, \mathfrak{g}_2^{\bot}\ .
\ee
This induces a decomposition of the space of 2-forms in the following
irreducible $G_2$-modules, 
\be981
\Lambda^2 \, = \, \Lambda^2_7 \, \oplus \, \Lambda^2_{14} \, ,
\ee
where
\[
\ba{rcl}
\Lambda^2_7 & = & \{u\haken\varphi | u\in TX_7 \} 
    =  \{ \alpha \in \Lambda^2 \, | 
           \ast(\varphi \wedge \alpha)-2\alpha=0 \} \\
\Lambda^2_{14} & = & \{ \alpha \in \Lambda^2 \, | 
           \ast(\varphi \wedge \alpha)+\alpha=0 \} \cong \mathfrak{g}_2
\ea
\]
and we introduced the abbreviation $u \haken \varphi \equiv u^{m}
\varphi_{mnp}$ and $\varphi$ denotes the $G_2$-invariant 3-index
tensor, which is expressed as fermionic bi-linear in (\ref{622}).  The
operator $\ast(\varphi \wedge \alpha)$ splits therefore the 2-forms
correspondingly to the eigenvalues $2$ and $-1$.  These relations
serve us to define the orthogonal projections $\mathcal{P}_k$ onto the
$k$-dimensional spaces:
\bea930
\mathcal{P}_7(\alpha) & = & {1 \over 3} \, 
              (\alpha + \ast(\varphi \wedge \alpha)) 
    =  {1 \over 3} \, ( \alpha + {1 \over 2} \alpha\haken\psi) \ ,\\
\mathcal{P}_{14}(\alpha)&  = & {1 \over 3} \, 
              (2\alpha - \ast(\varphi \wedge \alpha))
   =  {2 \over 3} \, ( \alpha - {1 \over 4} \alpha\haken\psi) 
\eea
where $\psi = \ast \varphi$.  To be concrete, the $G_2$-singlet spinor
satisfies the condition
\[
({\cal P}_{14})_{ \ ab}^{cd}\, \gamma_{cd} \, \theta_0 =
{2 \over 3} \Big( \1^{cd}_{\ \ ab} - {1 \over 4} \psi^{cd}_{\ \ ab} \Big) 
\gamma_{cd} \, \theta_0 =0 \ .
\]
This constraint is equivalent to the condition
\be552
\gamma_{ab} \theta_0 = i \varphi_{abc} \gamma^c \theta_0 
\ee
which gives after multiplication with $\gamma$-matrices
\be512
\ba{rcl}
\gamma_{abc} \theta_0 &=& \Big( i \varphi_{abc} + \psi_{abcd} \gamma^d \Big)
  \, \theta_0 \ , \\
\gamma_{abcd} \theta_0 &=& \Big( - \psi_{abcd} - 4 i \varphi_{[abc} 
        \gamma_{d]} \Big) \theta_0 \ .
\ea
\ee
Since it is a normalized spinor and due to the properties of the 7-d
$\gamma$-matrices (yielding especially $\theta_0^T \gamma_a \theta_0
=0$), we get from these relations the following bi-linears
\be622
\ba{rcl}
1&=& \theta_0^T \theta_0  \ ,  \\[1mm]
i\, \varphi_{abc} &=&  \theta_0^T \gamma_{abc} \theta_0  \ , \\[1mm] 
- \psi_{abcd} &=&  \theta_0^T \gamma_{abcd} \theta_0  \ , \\[1mm]
i\, \epsilon_{abcdmnp} &=&  \theta_0^T\gamma_{abcdmnp} \theta_0  \ .
\ea
\ee
All other bi-linears vanish.  These identities can now be used to
derive projectors for the complex 7-spinor in (\ref{726})
\[
\ba{rcl}
\gamma_a\theta &=& {e^Z \over \sqrt{2}} (\gamma_a + v_a 
       + i\varphi_{abc}v^b\gamma^c)\theta_0 \ , \\
\gamma_{ab}\theta &=& {e^Z  \over \sqrt{2}}(i\varphi_{abc}\gamma^c
       +i\varphi_{abc}v^c + \psi_{abcd}v^c\gamma^d
       -2v_{[a}\gamma_{b]})\theta_0\ , \\
\gamma_{abc}\theta &=& {e^Z  \over \sqrt{2}}(i\varphi_{abc} 
     + \psi_{abcd}\gamma^d +3iv_{[a}\varphi_{bc]d}\gamma^d
     - \psi_{abcd}v^d -4i\varphi_{[abc}\gamma_{d]}v^d)\theta_0\ , \\
\gamma_{abcd}\theta &=& {e^Z  \over \sqrt{2}} (-\psi_{abcd} 
     - 4i \varphi_{[abc}\gamma_{d]} - 5\psi_{[abcd}\gamma_{e]}v^e
     - 4i v_{[a}\varphi_{bcd]} - 4v_{[a}\psi_{bcd]e}\gamma^e )\theta_0\ , \\
\gamma_{abcde}\theta &=& {e^Z  \over \sqrt{2}} (-5\psi_{[abcd}\gamma_{e]}
     -i\varepsilon_{abcdefg}\gamma^g v^f -5 v_{[a} \psi_{bcde]}
     -20i v_{[a} \varphi_{bcd}\gamma_{e]} )\theta_0\ , \\
\gamma_{abcdef}\theta &=& {e^Z  \over \sqrt{2}} 
   (-i\varepsilon_{abcdefg}\gamma^g
     + \varepsilon_{abcdefg}v_h\gamma_j\varphi^{ghj} 
     -i \varepsilon_{abcdefg}v^g )\, \theta_0 \ .
\ea
\]
Using complex notation, we can introduce the following two sets of bi-linears 
[$\hat{\theta}^{\dagger}=(\hat{\theta}^{\ast})^T$]:
\[
\Omega_{a_1 \cdots a_k } \, \equiv \, 
     {\theta}^{\dagger}\gamma_{a_1 \cdots a_k} {\theta}
\qquad \mbox{and}\qquad
\tilde \Omega_{a_1 \cdots a_k} \, \equiv \, 
     {\theta}^T\gamma_{a_1 \cdots a_k} {\theta}
\]
and we define the associated $k$-forms by
\[
\Omega^k \, \equiv \, 
     {1 \over k!} \Omega_{a_1 \cdots a_k} e^{a_1 \cdots a_k}
\qquad \mbox{and}\qquad
\tilde \Omega^k \, \equiv \, 
     {1 \over k!} \tilde\Omega_{a_1 \cdots a_k} e^{a_1 \cdots a_k} \, .
\]
Thus, we obtain the following forms (cp.\ also \cite{110})
\be910
\ba{rcl}
\Omega^0 & = & e^{2 \, {\rm Re}(Z)} \ , \\
\Omega^1 & = & e^{2\, {\rm Re}(Z)} \, v \ , \\
\Omega^2 & = & i \, e^{ 2\, {\rm Re}(Z)} \, v\haken\varphi \, 
                = \, i \, e^{ 2\, {\rm Re}(Z)} \, \omega \ , \\
\Omega^3 & = & i \, e^{ 2\, {\rm Re}(Z)} \, 
    \Big[v\wedge ( v\haken\varphi)\Big]
        \, = \, i \, e^{ 2\, {\rm Re}(Z)} \, v \wedge \omega \ ,\\ 
\tilde\Omega^3 & = & i \, e^{ 2\, {\rm Re}(Z)} 
   \Big[ e^{2i\, {\rm Im}(Z)} \Big( \varphi \, - \, v\wedge\omega 
            \, - \, i \, v\haken\psi \Big) \Big] \, 
   = i \, e^{ 2\, {\rm Re}(Z)} \, \Omega^{(3,0)} \ , \\
\tilde \Omega^4 & = &  e^{2 {\rm Re}(Z)} \, \Big[ v \wedge (v\haken\psi) 
                    - i \, v \wedge \varphi \Big] =
-i \, e^{ 2\, {\rm Re}(Z)} v \wedge \, \bar \Omega^{(0,3)} \ , \\
\Omega^4 & = & - e^{2 {\rm Re}(Z)} \, 
              \Big[ \psi - v \wedge (v\haken\psi)\Big] =
- {1 \over 2}  e^{2\, {\rm Re}(Z)} \, \omega \wedge \omega\ .
\ea
\ee
The associated 2-form to the almost complex structure on $X_6$ is
$\omega$ and with the projectors ${1 \over 2} ( \1 \pm i \omega)$ we
can introduce (anti) holomorphic indices so that $\Omega^{(3,0)}$ can
be identified as the holomorphic $(3,0)$-form on $X_6$.  There exists
a topological reduction from a $G_2$-structure to a $SU(3)$-structure
(even to a $SU(2)$-structure).  The difficulties arise by formulating
the geometrical reduction. Using the vector $v$ one is  able to
formulate an explicit embedding of the given $SU(3)$-structure in the
$G_2$-structure to obtain such a geometrical reduction:
\be920
\ba{rcl}
\varphi & = & {\rm Re}( e^{-2 i\, {\rm Im}(Z)} \, 
\Omega^{(3,0)}) + v\wedge \omega \, =  \chi_+ + \, v \wedge \omega \ ,
\\
\psi & = & {\rm Im}( e^{-2i\, {\rm Im}(Z)} \, \Omega^{(3,0)})\wedge v \, 
           + \, {1 \over 2} \omega^2 \, =\,  
\chi_- \wedge v \, + \, {1 \over 2} \omega^2
\ea
\ee
with the compatibility relations
\bea945
e^{- 2i\,  {\rm Im}(Z)} \, \Omega^{(3,0)} \wedge \omega & = & 
(\chi_+ \, + \, i \, \chi_-) \wedge \omega
            \, = \, 0 \, , \\
\chi_+ \, \wedge \, \chi_- & = & {2 \over 3} \, \omega^3\ .
\eea
Since the phase factor, coming form ${\rm Im}(Z)$, does not play any
role in the following and we will set it to zero.  Before we will use these
relations for the discussion of the Killing spinor equations, we have
first to discuss the deformation of the geometry due to the fluxes.


\section{Back reaction onto the geometry}
\resetcounter


Due to gravitational back reaction the fluxes deform the geometry, not
only of the internal but also of the external space. This is related
to the fact that the 4- as well as the 7-d spinor is not anymore
covariantly constant. There is also an ongoing discussion
on this subject in the mathematical literature, see e.g.\ 
\cite{370, 580, 350, 510, 620}.

Consider first the 7-d internal space. The spinor is singlet under a
subgroup of the structure group and hence satisfies certain projector
relations; see previous section.  This can be used to re-write all
flux terms in the Killing spinor equation as con-torsion terms
entering the covariant derivative as follows
\[
\tilde \nabla_a \theta \equiv ( \nabla_a - {1 \over 4} \tau^{bc}_a
\gamma_{bc} ) \theta = 0 \ .
\]
{From} the symmetry it follows that it has $7 \times 21 = 7 \times
(7+14)$ components, but if $\theta$ is a $G_2$-singlet the ${\bf 14}$
drops out and hence $\tau \in \Lambda_1 \otimes {\mathfrak
g}_2^{\perp}$, where $\Lambda_1$ is the space of 1-forms and
${\mathfrak g}_2^\perp$ is defined by the Lie-algebra relation ${\mathfrak
g}_2^\perp \oplus {\mathfrak g}_2 = \mathfrak{so}(7)$. These
components decompose as
\[
{\bf 49} = {\bf 1 + 7 + 14 +27} = \tau_1 + \tau_7 + \tau_{14} + \tau_{27}
\]
where $\tau_i$ are called $G_2$-structures.  They are related to
differential forms that can be obtained from $d \varphi$ and
$d \psi$ as follows
\be093
\ba{rcl}
d\varphi & \in & \Lambda^4 \, = \, \Lambda^4_1 \, \oplus \, \Lambda^4_7
\, \oplus \, \Lambda^4_{27}  \ , \\
d \psi & \in & \, \Lambda^5 \, = \, \Lambda^5_7 \, \oplus \, 
\Lambda^5_{14}  \ ,
\ea
\ee
where the ${\bf 7}$ in $\Lambda^4_7$ is the same as in $\Lambda^5_7$
up to a multiple.  Note, since the spinor is not covariantly constant,
neither can be the differential form $\varphi$ and $\psi$. The different
components in the differentials $d\varphi$ can be obtained by using the
following projectors
\be971
\ba{rcl}
\mathcal{P}_{1}(\beta) &=&{1 \over 4!}
\psi \haken \beta \quad , \\
\mathcal{P}^4_{7}(\beta) &=&-{1 \over 3!}  \varphi \haken \beta , \\
\mathcal{P}_{27}(\beta)_{ab} &=& {1 \over 3!}(\beta_{cde\{a}\psi_{b\}}
{}^{cde})_0
\ea
\ee
where $\beta$ denotes a 4-form and in $( \cdot )_0$ we removed the
trace.  Let us summarize:
\be972
\ba{rcl}
\tau^{(1)} &\longleftrightarrow&  \psi \haken d\varphi \quad , \\
\tau^{(7)} &\longleftrightarrow&  \varphi \haken d\varphi \ , \\
\tau^{(14)} &\longleftrightarrow& 
              \ast d\psi - {1 \over 4} (\ast d\psi) \haken\psi \ , \\
\tau^{(27)} &\longleftrightarrow& 
              (d\varphi_{cde\{a}\psi_{b\}}{}^{cde})_0 \ .
\ea
\ee
With these definitions $\tau_{14}$ and $\tau_{27}$ have to satisfy:
$\varphi_3 \wedge \Lambda_{27}^3= \varphi_3 \wedge \tau_{14} =0$.
In the case of two 7-spinors, which combine to an SU(3)-singlet
spinor, these $G_2$-structures decompose into SU(3)-structures, which
consist of five components again related to the differential forms: $d
\omega$ and $d \Omega$, where the 2-form $\omega$ and 3-form $\Omega$
were introduced in (\ref{910}).  We refer here to literature for more
details \cite{340}. Note, the Killing spinor is invariant under the
$G$-structures group and therefore, the more spinors we have the less
is the $G$-structure group.

Let us end this section with the back reaction onto the external
space.  As we said at the beginning, we are interested only in the
case where this gives at most a cosmological constant yielding a 4-d
anti deSitter vacuum. As special case we recover of course the flat
space vacuum. In supergravity, the superpotential appears as mass term
for the gravitino and this means, that the corresponding Killing
spinor cannot be covariantly constant. Therefore, we introduce the
superpotential by assuming that our 4-d spinors solve the equation
\be811
\nabla_\mu \epsilon^i \  \sim \ \hat \gamma_\mu \, 
 e^{K/2}\, ( W_1^{ij} + i \hat \gamma^5 \,W_2^{ij} ) 
\, \epsilon_j \ 
\ee
where we took also into account a non-trivial K\"ahler potential $K$.
If there is only a single spinor as in eq.\ (\ref{251}), this equation
simplifies to
\[
\nabla_\mu \epsilon ~ \sim ~ 
\hat \gamma_\mu   e^{K/2} \, (W_1 + i \, \hat \gamma^5 \, W_2 ) \, \epsilon \ .
\]
If $\epsilon$ is a Weyl spinor, as in eq.\ (\ref{010}), this equation
becomes $\nabla_\mu \epsilon = \hat \gamma_\mu \bar W \epsilon^\star$
with the complex superpotential
\be252
W = W_1 +i\, W_2 \ .
\ee
Therefore, in the 11-d spinor equations, we introduce the
superpotential by assuming that 11-d spinor satisfies the
equation
\be152
\Big[\nabla_\mu \otimes \1 \Big] \eta = (\hat \gamma_\mu \otimes 
\1) \tilde \eta
\qquad {\rm with:} \qquad
\tilde \eta = \left\{ \ba{ll}  e^{K/2} \,
[( W_1  + i \, \hat \gamma^5 W_2) \otimes \1] \; \eta  \ & {\rm M} \\
 e^{K/2} \, W \epsilon \otimes \theta^\star + cc & {\rm W} \ea \right.
\ee
where M/W refers to a 4-d Majorana or Weyl spinor $\epsilon$.  We do
not consider here the case where the superpotential is matrix valued.


\section{BPS constraints}
\resetcounter


With the superpotential as introduced before, equation (\ref{112}) becomes
\be827
0 = \tilde \eta
+ \Big[\hat \gamma^5 \otimes \Big( {1 \over 2} \partial A + 
{i m \over 36} \Big)
+{1 \over 144} e^{-3 A}\, (\1 \otimes F) \Big] \eta \ .
\ee
and if 
\be927
\hat \eta = e^{- {A\over 2} } \eta
\ee
equation (\ref{711}) yields
\be553
0=  \1 \otimes
\Big( \nabla_a^{(h)} +{i \, m \over 48} \Big)\hat \eta
- i \,\hat \gamma^5 \hat \gamma_a \,  e^{-{A \over 2}} \tilde \eta 
  - { 1 \over 12} \,  e^{-3 A}\, \hat \gamma^5 \otimes F_a  \hat \eta \ .
\ee
We will now derive the constraints imposed by these two equations for
the two cases: $(i)$ that the 4-d spinor is Majorana or $(ii)$ that we
have two Weyl spinors of opposite chirality. This in turn implies that
we consider the cases that on $X_7$ is a single or two spinors.


\subsection{Manifolds with a single 7-spinor}


In this case, which has been discussed also in \cite{130}, we can use
the relations (\ref{512}) and find
\[
\ba{rcl}
F \theta_0 \equiv F_{abcd} \gamma^{abcd} \theta_0 &=&
\Big( - F_{abcd} \psi^{abcd} -4i F_{abcd} \varphi^{abc} \gamma^d \Big)
\theta_0 \ , \\
F_a \theta_0 \equiv F_{abcd} \gamma^{bcd} \theta_0 &=& 
\Big(i\, F_{abcd} \varphi^{bcd}
+ F_{abcd} \psi^{bcde} \gamma_e \Big)\, \theta_0 \ .
\ea
\]
Next, the internal part of the 4-form field strength has $35$
components, that decompose under $G_2$ as {\bf 35} $\rightarrow$ {\bf
1 + 7 + 27} with
\be881
F_{abcd} ={1 \over 7} \, \F \, \psi_{abcd} 
          + \, {\cal F}^{(7)}_{[a} \, \varphi_{bcd]}
          - 2 \, {\cal F}^{(27)}_{e[a} \psi^e_{\ bcd]} 
\ee
or
\be882
\ba{rcl}
\F &=& {1 \over 4!} \, F_{abcd} \psi^{abcd} \  , \\
\FF_a &=& {1 \over 3!} \, F_{abcd} \varphi^{bcd} \  ,\\ \label{364}
\FFF_{ab} - {4\over 7} \F \, \delta_{ab} &=& {1 \over 3!} \, 
           F_{cde\{a} \psi_{b\}}^{\ \ cde} \ .
\ea
\ee
Therefore, we can also write
\be241
\ba{rcl}
F \theta_0 &=& -4!\Big[ \F +i \FF_a \gamma^a \Big]\, \theta_0 \ , \\
F_a \theta_0 &=&  \Big[3! \,i\, \FF_a + \Big(F_{cde [a} \psi_{b]}^{\ \ cde}
+F_{cde \{a} \psi_{b\}}^{\ \ cde}\Big) \gamma^b \Big]\, \theta_0 \\
&=&
 3!\Big[i\, \FF_a + \Big(\varphi_{ab}^{\ \ c} \FF_c + \FFF_{ab} -{4 \over 7}
\F \delta_{ab} \Big) \gamma^b \Big]\, \theta_0 \ .
\ea
\ee
Note, by contracting with $\varphi$, one can verify that: $F_{cde[a}
\psi_{b]}^{\ \ cde} \sim \varphi_{ab}^{\ \ e} {\FF}_e$.  With these
relations it is now straightforward to solve the Killing spinor
equations. Because the 11-d spinor $\eta$ is Majorana, also the 7-spinor
$\theta$ as well as the 4-spinor $\epsilon$ have to be Majorana and as
consequence, terms ${\cal O}(\epsilon)$ and ${\cal O}(\hat \gamma^5
\epsilon)$ have to vanish separately.  We find the equations
\[
\ba{l}
0 = \Big( i  e^{K/2}\, W_2 + {1 \over 2} \partial A + 
{i m \over 36} \Big) \, \theta_0 
= \Big( i  e^{K/2}\, W_2 + {1 \over 2} \gamma^a \partial_a A + {i m \over 36} \
\Big) \, \theta_0  \ , \\
0 = \Big(  e^{K/2}\, W_1 +{1 \over 144}e^{-3A}  F \Big) \theta_0 
= \Big( e^{K/2}\, W_1 - {1 \over 6}e^{-3A} [\F +i 
\FF_a \gamma^a]  \Big)\, \theta_0 \ .
\ea
\]
Since terms ${\cal O}(\theta_0)$ and ${\cal O}(\gamma^a
\theta_0)$ cannot cancel, we get finally as solution
\be249
A = const. \ ,  \qquad  m  = - 36\,   e^{K/2}\, W_2 \ 
\qquad {\cal F}^{(7)}_a = 0\ , \qquad
{e^{-3A} \over 6}\F =  e^{K/2}\, W_1 \ .
\ee
The remaining differential equation (\ref{553}) can be solved
in the same way. With eq.\ (\ref{241}) and $\FF_a = 0$ we get the equations
\[
\ba{rcl}
0&=&\Big( \nabla_a - i \Big[  e^{K/2}\, 
W_2 - {m \over 48} \Big] \gamma_a \Big) 
\, e^Z \, \theta_0 =
\Big( \nabla_a + i \, {7 \, m \over 144} \, \gamma_a \Big) 
\, e^Z \, \theta_0 \ ,\\
0&=& \Big(  e^{K/2}\, W_1 \gamma_a + 
{1\over 12}\,e^{-3A}\, F_a \Big) \theta_0 =
\Big(  e^{K/2}\, W_1 \delta_{ab}  + {1 \over 2}\,e^{-3A}\,\Big[  
 \FFF_{ab}  -{4\over 7} \F \delta_{ab}\Big]\Big)  \gamma^b \theta_0  \ .
\ea
\]
Using (\ref{249}) we find from the second equation
\be768
W_1=\F = \FFF_{ab} = 0 
\ee
and thus all internal 4-form components have to vanish
\be311
F_{abcd} = 0  \ .
\ee
The other equation gives now a differential equations for the spinor
$e^Z \theta_0$.  The factor $e^Z$ does not contain any
$\gamma$-matrices\footnote{Recall, as a $G_2$ singlet $\theta$ has
only one component, which we normalized by allowing a non-trivial
factor $e^Z$.}  and if we contract this equation with $\theta_0^T
e^{-Z}$ and use the fact that $\theta^T_0 \theta_0 = 1$ (and
therefore $\theta^T_0 \nabla_a \theta_0 = 0$) we find 
\[
\partial_a Z =0
\]
and this constant factor can be dropped from our analysis.  The
differential equations for $\theta_0$ fixes the 7-manifold to have a
weak $G_2$ holonomy.  In fact, after taking into account the
vielbeine, this gives the known set of first order differential
equations for the spin connection 1-form $\omega^{ab}$
\be321
\omega^{ab}  \varphi_{abc} = { 7\over 36} \, m \, e_b
\ee
where $m$ was the Freud-Rubin parameter [note $\omega$ is here the
spin connection and should not be confused with the associated
2-form introduced before]. 
 
Using the differential equation for the 7-spinor, it is
straightforward to verify that
\[
d\varphi =-{7 \, m \over 18} \, \psi \quad ,\qquad d \psi = 0
\]
and therefore only $\tau^{(1)}$ is non-zero.

To conclude this case, we have verified a 7-manifold of weak $G_2$
holonomy, which is Einstein and the cosmological constant is given by
the Freud-Rubin parameter \cite{150,130}. Note also, that the theory
of weak holonomy, $G_2$-structures forces $m$ to be a real
constant. The holonomy group with respect to the Levi-Civita
connection reduces in this case {\em not} to the group $G_2$ and is
therefore of generic type.  Only in case of $m=0$ we get back to the
special manifolds having precisely holonomy group $G_2$. We found that
in this case all internal flux components have to vanish.  The 4-d
superpotential is only given by the Freud-Rubin parameter, ie.
\be883
 W \ \sim \ i \int_{X_7} {^{\star} F}
\ee
which fixes the overall size of the 7-manifold.  In the limit of flat
4-d Minkowski vacuum, the Freud-Rubin parameter has to vanish and we
get back to the Ricci-flat $G_2$-holonomy manifold. In order to allow
for non-trivial fluxes we have therefore to consider the case with two
spinors on $X_7$ or equivalently to use 4-d Weyl spinors.


\subsection{Manifolds with two 7-spinors}


The case before was rather trivial, as next step we consider the more
general spinor as given in (\ref{010}), which comprises of two 4- and
7-spinors and we shall solve again the equations (\ref{827}) and
(\ref{553}). But now, the complex 4-spinor is chiral and we choose
\[
\gamma^5 \epsilon = \epsilon \quad , \qquad \gamma^5 \epsilon^\star
= - \epsilon^\star \ .
\]
Due to the opposite chirality, terms with $\epsilon$ and
$\epsilon^\star$ are independent and the term ${\cal
O}(\epsilon)$ in eq.\ (\ref{827}) becomes
\be660
0  =  e^{K/2}\, W \theta^\star + \Big(
{i m \over 36}  +{1 \over 2} \partial A + {1 \over 144}
\,e^{-3A}\, F \Big) \theta  \ .
\ee
The equation coming from the terms ${\cal O}(\theta^\star)$ do not
contain new information, but gives just the complex conjugate of this
equation.  Now, with the relations from section 3  we can collect all terms
${\cal O}(\theta_0)$ and ${\cal O}(\gamma^a \theta_0)$ which have to
vanish separately and find
\bea661
0   &= &  e^{K/2}\, W + {i m \over 36}  +{1 \over 2 } 
v^a \partial_a A - {1 \over 6 }\,e^{-3A}
\Big[\F + i v^a \FF_a \Big]  \\ \label{228}
0&=& \Big[ -  e^{K/2}\, W + {im \over 36} \Big] v_a + { 1\over 2}
\Big[\delta_{ab} + i \varphi_{abc} v^c \Big] \, \partial^b A \\ \nonumber
&&-  {1 \over 6}\,e^{-3A} \Big[ \F v_a  - i \FF_a + {1 \over 3}
 F_{bcd\{a} \psi_{e\}}{}^{bcd} v_e
\Big] \ .
\eea
And separating the real and imaginary part, we get four equations
\bea150
0&=&  e^{K/2}\, W_1 - {1 \over 6}\,e^{-3A}\, \F + 
{1 \over 2} v^a \partial_a A  \ , \\
0&=&  e^{K/2}\, W_2 + {m \over 36} - 
{1 \over 6}\,e^{-3A}\, \FF_k v^k\ , \\ \label{151}
0&=& [ -  e^{K/2}\, W_1  + {1 \over 42} \,e^{-3A}\, \F ]\, v_a + 
{1 \over 2}\partial_a A 
      - {1 \over 3} \,e^{-3A}\, \FFF_{ak} v^k  \ , \\ 
0&=& \big[ -  e^{K/2}\, W_2 + {m \over 36}\big] \, 
v_a +{1 \over 6}\,e^{-3A}\, \FF_a 
      - {1 \over 2}\varphi_{abc}v^b \partial^c A   \ .
\eea
If we contract the last two equations with the vector $v^a$ and 
add/subtract the results to the first two equations we infer
\be524
\ba{rcl}
W= W_1 + i \, W_2 
&=& {1 \over 6} \Big[ {4 \over 7} \F - v^a \FFF_{ab} v^b + i v^a \FF_a
  \Big] \\
v^a \partial_a e^{3A} &=&  {3 \over 7} \F + v^a \FFF_{ab} v^b 
\\
m &=&0 
\ea
\ee
We set $K = - 6 A$, i.e.\ identified the K\"ahler
potential\footnote{At least for specific K\"ahler gauge this is
possible and appeared also in the supergravity solutions \cite{190}.}
with the warp factor, which yields a holomorphic superpotential, see
eq.\ (\ref{339}) below. In the discussion we will re-write these
expressions in terms of quantities on the 6-manifold $X_6$. For the
other flux components we find
\bea525
 (\delta_a^{\ b} - v_a v^b ) \FF_b & = &   
 \varphi_{abc} v^b \partial^c e^{3A} \ =\ 2 \, \varphi_a^{\ bc} v_b 
       \, \FFF_{cd} v^d 
\ , \\   \label{651}
2\, \FFF_{ab} v^b & = & \Big[- {3 \over 7}  \F +  v^c \FFF_{cb} v^b
\, \Big] v_a +  \partial_a e^{3A}
\eea
[the flux components were introduced in (\ref{881})].

Finally, we have to investigate the internal part (\ref{553}) of the
variation giving
\bea722
\nabla_a^{(h)} \hat \theta =  e^{K/2}\, W \gamma_a \, \theta^\star \, 
  +  {1\over 12}\,e^{-3A} \, F_a  \, \hat \theta 
\eea
with $\hat \theta = {e^{-{A \over 2 } + Z} \over \sqrt{2}} (\1 + v)
\theta_0$.  On the rhs of this equation we can use our previous
formulae and write
\[
F_a \hat \theta = {e^{Z - {A \over 2}} \over \sqrt{2}}
( F_a{}^1 + F_a{}^7 +  F_a{}^{27})\theta_0
\]
where  
\[
\ba{rcl}
\gamma_a \hat \theta^\star & =& 
 {1 \over \sqrt{2}}\, e^{Z^\star}\,  (\gamma_a - v_a 
       - i\, \varphi_{abc}v^b\gamma^c)\theta_0 \ , \\
F_a{}^1 \theta_0 & = & {24 
\over 7  } \F (v_a - \gamma_a) \theta_0 \ , \\
F_a{}^7 \theta_0 & = & 3 \, \Big[ 2i \FF_a 
   - \varphi_{ab}^{\ \ c} (v^b - \gamma^b) \, \FF_c
   -i\, (v_a\gamma^b+\gamma_a v^b) \, \FF_b \Big]\theta_0 \ , \\
F_a{}^{27}\theta_0 & = & 6 \, \Big[  
          -\FFF_{ab}(v^b -  \gamma^b)   + i\varphi_{ad}^{\ \ b}\, 
           (\gamma^c v^d + \gamma^d v^c)\, \FFF_{bc}\, \Big]\, \theta_0 \ .
\ea
\]
With these expressions we can now calculate the $G$-structures
as introduced in (\ref{093}) and find
\be761
\ba{rcl}
\tau^{(1)}  & \longleftrightarrow &  W_2 \ , \\
\tau^{(7)}_a \, &\longleftrightarrow& \, 48\, W_1 \, v_a
                        -{24 \over 7} \, {\cal F}^{(1)} \, v_a
          + \,  {3 \over 2} \, \varphi_{a}{}^{bc} v_b \FF_c
+ 27 \FFF_{ab} v^b \ .
\ea
\ee
We used here the constraint
\be771
0 \, = \, d [{-A  + 2 {\rm Re}(Z)}]  
\ee
which one can derive from
\[
\partial_a \| \hat \theta \|
\equiv \partial_a [\hat \theta^{\dagger} \hat \theta ]
  = 
    e^{-A  + 2 {\rm Re}(Z)} \, \partial_a [{-A  + 2\, {\rm Re}(Z)}] \, = \, 
   (\nabla_a \hat{\theta})^T \hat{\theta}^{\ast} +
       \hat{\theta}^T \nabla_a(\hat{\theta}^{\ast}) = 0\\
\]

To make the set of equations complete, we have to derive the
differential equations obeyed by the vector field $v$, which is
straightforward if we use again the differential equation for the
spinor. To simplify the notation and for later convenience we use here
the 2-form $\omega$ which was introduced in (\ref{920}) and
find for the covariant derivative
\be120
\ba{rcl}
\nabla_{m} v_{n} &=&  
- {1 \over 12} e^{-3A -2{\rm Re}(Z)}\, \theta^+ [ F_m , \gamma_n] \theta \\
&=&   {1 \over 12} 
e^{-3A} F_{mbcd} \omega^{bc} \omega^{d}_{\ n}
\ea
\ee
recall $\omega_{ab} = \varphi_{abc}v^c$. Note, $v^n \nabla_m v_n = 0$,
which is consistent with $|v|^2 =1$. Using the decomposition
(\ref{881}) one finds for the (anti)symmetrized components
\bea121
\nabla_{[m} v_{n]} &=& \Big(\delta_{[m}^{\ a} \delta_{n]}^{\ b}
     +{1\over2}\psi_{mn}{}^{ab}\Big) \FFF_{ac}v^c v_b 
              +{1\over4}\varphi_{mn}{}^a(\delta_a^{\ b} -v_a v^b) \FF_b
         \ , \\[3mm]
\nabla_{\{m} v_{n\}} & = &  
      - {2 \over 7} (\delta_{mn} - v_{\{m} v_{n\}})\, \F
    -{1\over 2} v_{\{m}\varphi_{n\}}{}^{ab}v_a\FF_b  \nonumber \\ 
&&  +{1 \over 2}\Big(
         \delta_m^{\ a} \delta_n^{\ b} + \omega_m^{\ a} \omega_n^{\ b}\Big)
           \FFF_{ab}
      -{1 \over 2}\delta_{mn}\FFF_{ab} v^a v^b  
 \eea
The first term in the anti-symmetric part is the projector onto the
${\bf 7}$, see (\ref{930}), and by contracting with $\varphi$ and
employing eqs.\ (\ref{525}) and (\ref{651}), one can verify that
\[
d(e^{3A}v)=0 \ .
\]
Note, in comparison with \cite{110} our internal metric is conformally
rescaled.


\section{Discussion}

\resetcounter


Let us now summarize and discuss our results. We considered flux
compactifications of M-theory on a 7-manifold $X_7$, where we
distinguished between the number of (real) spinors.  The case with a
single 7-d spinor was rather trivial: all (internal) fluxes had to
vanish and the Freud-Rubin parameter deformed $X_7$ to a manifold with
weak $G_2$-holonomy. Let us concentrate here on the situation with two
7-d spinors with the constraints derived in the previous subsection.
In contrast to the single spinor case, the Freud-Rubin parameter has
to vanish in this case. The no-where vanishing vector field $v$, which
was built as fermionic bi-linear, can be used to foliate $X_7$ by a
6-d manifold $X_6$ and for the discussion it is useful to project the
fluxes onto $X_6$, which in turn is equivalent to the decomposition
under SU(3). So we define
\be629
H_{abc} = v^m F_{mabc}  \quad (\, H \equiv v \haken F \, )
 \quad , \qquad  G_{abcd} = F_{abcd}
\ee
where the indices $a,b, \ldots = 1, \ldots , 6$ are related here to
coordinates on $X_6$ and we denote the 4-form on $X_6$ by $G$.
Strictly speaking, this reduction is valid only as long as $v$ is not
fibered over $X_6$, which would result in a non-trivial connection $A$
and hence would result in a Chern-Simons term in $G$ of the form $A
\wedge H$. This is well known from Kaluza-Klein reduction, but since
this is not important for the discussion here, we will define $G$
without this Chern-Simons term.  Actually we are not doing the
dimensional reduction ($v$ is in general not Killing), we only project
the equations onto $X_6$ and distinguish between different components.
Now, counting the number of components and decomposing them under
SU(3) we find
\[
\ba{rcl}
{[ H ]} &=& {\bf 20} = {\bf 6 + \bar 6 +  3 + \bar 3  + 1 + \bar 1}  \\
{[ G ]} &=& {\bf 15} = {\bf 8 + 3 + \bar 3  + 1} 
\ea
\]
where we used the projector ${1\over 2} (\1 \pm i\, \omega)$ to
introduce (anti) holomorphic indices on $X_6$. 

Our main results are in eqs.\ (\ref{524}) -- (\ref{651}) combined with
eqs.\ (\ref{120}) and let us now discuss the BPS constraints on the
different (projected) components.  

The ${\bf 1 + \bar 1}$ of $H$ are the (3,0) and (0,3) part, which
fixes the superpotential as written in the form of
\be339
W = { i \over 36} \bar \Omega^{(0,3)} \haken H \ \sim\  
{1 \over 36} \int_{X_7} F \wedge \Omega^{(3,0)} \ .
\ee
This is obviously a holomorphic superpotential, which can be verified
{from} (\ref{524}) by contracting the last two equations in
(\ref{364}) with $v^a$.  It can also be obtained by contracting the
spinor equation (\ref{660}) by $\theta^T$ which gives the holomorphic
expression
\[
W\  \sim \ \theta^T F \theta \ .
\]
Note, the differential forms $\tilde \Omega^{k}$ as given in
(\ref{910}) are only non-zero for the 3- or its dual 4-form.

Similarly, one can verify that the ${\bf 3 + \bar 3}$ components of
$G$ have to vanish. They are given by the (3,1)-part of $G$ and we can
project onto them by contracting (\ref{660}) with $\theta^T \gamma_a$
which gives $0= \theta^T \gamma_a F \theta \sim G_{a}^{\ bcd}
\Omega_{bcd}$ on $X_6$, see again (\ref{910}). But we can also see
this by using (\ref{920}) to write
\[
\FF\big|_{X_6} \equiv -{1 \over 3!} \varphi \haken F\big|_{X_6} \  =\    
- {1 \over 6} (\chi_+) \haken G - {1 \over 2} \, \omega \haken H
= d(e^{3A}) \haken \omega
\]
where we used eqs.\ (\ref{525}) and (\ref{651}). Similarly, we get
\[
2 v \haken \FFF\big|_{X_6} \ = \ - {1 \over 6}
 (\chi_-) \haken G  + {1 \over 2} H \haken \omega^2 = d(e^{3A})\big|_{X_6} 
\ .
\]
These two equations imply now that
\be331
\Omega \haken G = 0 
\ee
and 
\be889
d e^{3A} \haken \omega = {1 \over 2}  \omega \haken H   \ .
\ee
Therefore, the ${\bf 3 + \bar 3}$ of $G$ has to vanish defining $G$ as
a (2,2)-form and the warp factor is fixed by the non-primitive
(2,1)-part of $H$ (i.e.\ its ${\bf 3 + \bar 3}$).

Thus, the only non-zero components are so far
\be092
{[ H ]} = {\bf 6 + \bar 6 +  3 + \bar 3 + 1 + \bar 1} \quad, \qquad
{[ G ]} = {\bf 8 + 1}  \ .
\ee
The ${\bf 1 + 8}$ of $G$ are the components of a (2,2)-form, which is
equivalent to the (1,1)-form obtained by $\omega \haken G$. The ${ \bf
1}$ is the component that fixes the gradient of the warp factor along
the vector $v$
\be669
{1 \over 2\, 4!} \, G^{(1)} \equiv {1 \over 144} \omega^2 \haken G  
=  \, \Big( {3 \over 7} \F + v^a \FFF_{ab} v^b \Big) =
 v^a \partial_a e^{3A}
\ee
which is obtained from (\ref{651}).  The ${\bf 8}$ is the (1,1) part
in $\FFF_{ab}$ and hence we can project onto it by\footnote{Note the
contraction of (anti) holomorphic indices with $\omega$ yield ``$\pm
i$'' so that this combination projects onto the (1,1)-part.}
\[
\Big[\delta_{c}^{\ a} \delta_{d}^{\ b} 
     + \omega_{c}^{\ a}  \omega_{d}^{\ b} \Big]\, \FFF_{ab} 
\]
and it appears in (\ref{120}) as the traceless symmetric component.

The remaining components of $\FFF_{ab}$, comprise a symmetric (2,0)
and (0,2) tensor and give the ${\bf 6 + \bar 6}$ that enter $H$.
These 12 components are related to the primitive (2,1)-part of $H$,
i.e.\ the components that satisfy: $H^{(2,1)} \wedge \omega = 0$.
They drop out in all our expressions and hence these
components are unconstrained.

We might also ask whether $X_6$ is a complex manifold, i.e.\ whether
$d\omega \in \Lambda^{(2,1)} \oplus \Lambda^{(1,2)}$ and $d\Omega \in
\Lambda^{(3,1)}$ when projected on $X_6$ \cite{381}.  These
requirements mean, that $d\omega$ has no (3,0) or (0,3) part and
similarly $d\Omega$ has no (2,2) part (see also \cite{330, 170}).
Using the differential equations for the spinor (\ref{722}), we find
however that
\[
\Omega^{(3,0)} \wedge d \omega \ \sim  \  W \, i \,( \Omega^{(3,0)} \wedge
\bar \Omega^{(0,3)})
\] 
and hence $X_6$ can only be complex if the superpotential vanishes.
In fact, it is straightforward to show that in this case also
$(d\Omega)^{(2,2)} = 0$.

But this expression implies also that we can express the
superpotential purely geometrically (without fluxes) by
\[
W \ \sim \  \int_{X_7} i v \wedge d\omega \wedge \Omega^{(3,0)}
\]
which has to be contrasted with the form derived in (\ref{339}). Since
our setup is only sensible to value of the superpotential in the
vacuum it is difficult to distinguish between both expressions. For
the case that $v$ is Killing, we can however compare our results with
the ones derived in \cite{330, 421, 180} and this suggest that we have
to add both expressions yielding
\be372
W  \ \sim \  \int_{X_7} (F + {i} v \wedge d \omega)\wedge \
\Omega^{(3,0)} \ .
\ee
Upon reduction to 10 dimensions, this superpotential contains only
fields that are common in all string theories and this proposal
based on a number of consistency checks (that it is U-dual to the type
IIB superpotential and anomaly-free on the heterotic side).


\section{Examples}
\resetcounter


So far, our discussion was general and let us now consider specific
 cases in more detail.

\medskip

\noindent
{\em Case (i): The vector $v$ is closed ($dv=0$)}

\nopagebreak
\noindent
If $v$ is closed, we can (at least locally) write $v = dz$, where the
coordinate $z$ can be regarded as the $11^{th}$ or $5^{th}$
coordinate. {From} (\ref{120})  and the BPS constraints (\ref{525}) we
obtain now
\[
v \wedge dA=0 \quad \rightarrow  \quad \FF_a = \FFF_{am} v^m = 0 \ .
\]
Therefore, only the non-vector components of $\FFF$ and $\F$ are
non-zero and in the notation of (\ref{092}), the ${\bf 3 + \bar 3}$
have to vanish.  It is natural to consider a reduction over $X_6$ to
obtain an effective 5-dimensional description.  Note, $X_6$ does not
need to be simple connected and one may also consider the case where
$X_6$ factorizes into different components with different
$z$-dependent warping. In any case, the resulting {5-d} solution is
not flat, but our 4-d external space represents a domain wall with $z$
as the transversal coordinate. Since the fluxes on $X_6$ are
non-trivial we get a potential for the 5-d theory, which agrees with
the real superpotential in \cite{160, 200}
\[
W_{5d} \sim {1 \over 144} \, G^{(1)} = {1 \over 144}
\, \int_{X_6} G \wedge \omega \ .
\]
Let us mention here a subtle issue. For the case discussed in this
paper (i.e.\ SU(3) structures), the 4-d superpotential $W$ does not
come from the 5-d potential, but represents in 5 dimensions a kinetic
term of two (axionic) scalars of a hyper multiplet\footnote{These are
the scalars coming from the (3,0) and (0,3) part of the 11-dimensional
3-form potential.}, which is non-zero in the vacuum and curves the
domain wall.  That kinetic terms of (axionic) scalars act effectively
as a potential can be understood from massive T-duality \cite{441,442}
(if one allows for a linear dependence of the $5^{th}$ direction) and
we expect that these flux compactifications are dual to the curved
domain walls discussed in \cite{440,450,430}.  The 5-d superpotential
$W_{5d}$ is instead compensated by the warp factor as in eq.\
(\ref{669}), which is the known (5-d) BPS equation (in proper
coordinates)
\[
\partial_z A(z) = - 3 \, W_{5d} \ .
\]
It is also instructive to discuss the fixing of the moduli.  If we
start with the scalars entering $W_{5d}$ and expand the 4-form $G$ as
well as $\omega$ in a complete basis of harmonic 4- and 2-forms,
resp. As result, the superpotential becomes $W_{5d} = p_I t^I =
e^{2\varphi} p_I X^I$, where $p_I$ are the coefficients from the
expansion of the 4-form $G$, $t^I$ are the K\"ahler class moduli,
which are rescalled by the volume scalar $e^{2\varphi}$. The resulting
fields $X^I=X^I(\phi^A)$ parameterize the vector multiplet moduli
space ${\cal M}_V$ of 5-d (gauged) supergravity and {\em all} of them
are fixed in the vacuum. In fact their value can be determined
explicitly using the so-called attractor mechanism\footnote{It was
originally introduced to calculate the black hole entropy, but can
also be used in gauged supergravity to determine the fixed point
values of vector multiplet moduli. A number of explicit solutions of
these algebraic equations are known in 5 but also 4 dimensions
\cite{290, 200, 210, 190}.} \cite{310}. The (5-d) vacuum is namely
defined by the equations: ${\partial \over \partial \phi^A} W_{5d} =
p_I {\partial X^I \over \partial \phi^A}=0$ and since ${\partial X^I
\over \partial \phi^A}$ is a tangent vector on the moduli space ${\cal
M}_V$, the vacuum is given by the point where the flux vector $p_I$ is
normal on the manifold ${\cal M}_V$, see \cite{200} for more details.
In a generic situation this is a point on ${\cal M}_V$ and thus {\em
all} vector moduli are fixed by these fluxes.  Note, the volume scalar
gives a run-away behavior of the potential, which results into a
power-like behavior of the warp factor in the coordinate $z$
(especially these models do not give rise to an exponential warping)
and consequently the supergravity solution exhibits a naked
singularity. 

How about the complex structure moduli? They are related to scalars in
hyper multiplet, which parameterize a quaternionic manifold and they
enter the 5-d superpotential only if it is SU(2)-valued (coming from
the momentum maps), see e.g.\ \cite{560, 590}. This is beyond our
spinor ansatz, but we do have 3-form fluxes on $X_6$, which enter the
4-d superpotential and might lift the complex structure moduli space
nevertheless. In fact, for the 4-d superpotential as given in the form
(\ref{339}) we could repeat the procedure as done for $W_{5d}$ and the
corresponding 4-d attractor formalism will yield fixed point
values. But since $X_6$ is not a complex manifold for $W \neq 0$, the
expansion in harmonic forms becomes subtle and we do not know whether
this is a reliable procedure. On the other hand, the 4-d
superpotential corresponds in 5 dimensions to kinetic terms of scalar
fields in hypermultiplets. Because these kinetic terms do not vanish
in the vacuum and the quaternionic metric is not invariant under
generic deformations of the hyperscalar field, the complex structure
moduli space seemed to be lifted for all deformation that do not
represent isometries of the quaternionic space.

Setting $H=W=0$ our setup describes the model discussed in
\cite{290,200}. In this case $X_6$ is K\"ahler and it is
straightforward to introduce a brane which is a source for the flux
and corresponds to an M5-brane wrapping a holomorphic curve in $X_6$.
The worldvolume of the M5-brane can be identified by the 6-form:
$\ast(dA \wedge F)$. Since $dA \sim dz \sim v$, the 5-brane does not
wrap the $v$-direction, but the 2-cycle corresponding to the
(1,1)-form $\ast_6 G\ $ [note $G$ is (2,2) form containing the SU(3)
representations ${\bf 1 + 8}$].

\medskip

\noindent
{\em Case (ii): $v$ is Killing, i.e.\  $\nabla_{\{m}
v_{n\}} =0$}

\nopagebreak
\noindent
If both indices are on $X_6$, eq.\ (\ref{120}) implies that
$\nabla_{\{a } v_{b\}}$ is fixed by the components ${\bf 1 + 8}$ of
$G$ and if one index is along $v$, i.e.\ $v^m\nabla_{\{m } v_{b\}}$,
which is  $\omega_{bc} \partial^c A$. Since the $G^{(3,1)} = 0$
we conclude that $\nabla_{\{m } v_{n\}}=0$ implies
\[
G_{abcd} =0 \qquad, \qquad dA \big|_{X_6}  =0 \ .
\]
Note, if $G=0$ also $v^m \partial_m A=0$ and thus the warp factor has
to be constant, see (\ref{669}). Therefore, the only non-zero flux
components are the ${\bf 1+ \bar 1 + 6 + \bar 6}$ of $H$. If in
addition, the superpotential vanishes, only the primitive $H^{(2,1)}$
is non-zero and since $d \omega|_{X_6} \in \Lambda^{(2,1)}_0$ and $d
\Omega|_{X_6} =0$, $X_6$ is an Iwasawa manifold.  Because $v$ is a
Killing vector, we can make a dimensional reduction and these
manifolds have been discussed in string theory in more detail in
\cite{170}. Note, since $|v|^2 =1$, the 10-d dilaton as well as the
warp factor is constant for these string theory vacua.

\medskip

\noindent
{\em Case (iii): $\tilde v \equiv e^{\phi} v$ is Killing, i.e.\  $\nabla_{\{m}
v_{n\}} = - v_{\{n} \partial_{m\}} \phi$ }

\nopagebreak
\noindent
This case is closely related to the last case. Again we get
\[
G_{abcd} = 0 \ ,
\]
but now, all components of the $H$-field can be non-zero and the ${\bf 3 +
\bar 3}$ gives $d \phi$.  We find
\[
\phi = -3 A
\]
and note $\phi$ as well as $A$ do not depend on the coordinate along
the vector $v$, because from (\ref{669}) follows for $G=0$ that $v^m
\partial_m A = 0$.  If we now use this Killing vector $\tilde v$ for
dimensional reduction, the scalar field $\phi$ becomes up to a
rescaling the dilaton ($|\tilde v|^2 = e^{2\phi}$). Since all
components of the 4-form $G$ are zero, this case describes the common
sector of all string models and if the superpotential vanishes, we
obtain the vacuum already discussed in heterotic string models
\cite{530,170} (with trivial gauge fields; see \cite{570} for a recent
review).

\medskip

\noindent
{\em Case (iv): realization of MQCD}

\medskip

\noindent
Finally, we want to discuss the relation to the setup in
\cite{100,460,131}, where a 4-dimensional N=1 field theory was
realized on a M5-brane that wrap a 2-cycle in the internal space. In
10 dimensions this configuration corresponds to an intersection of two
NS5-branes, which are rotated by an SU(2) angle and D4-branes end on
both NS5-branes. In 11 dimension the M5-branes wrap a 2-cycle in a
6-dimensional space, which we identify with $X_6$ in our setup and
which means that the $11^{th}$ coordinate is {\em not} along $v$, but
inside of $X_6$. Of course, our notation might now be misleading,
since the $H$ field in our setup is not related to the $H$ appearing
after dimensional reduction to 10 dimensions along $v$.  Recall the
2-cycle wrapped by the M5-branes is identified by the 2-form $\ast(d A
\wedge F)$ and in the simplest case as assumed in \cite{100}, the warp
factor does not depend on the coordinate along $v$ (i.e.\ $v^m
\partial_m A =0$) and therefore $dA$ lies always inside $X_6$.  As
consequence, the 4-form $F$ has always a $v$ component and therefore
$G$ is zero and only the $H$ field is non-trivial. This is exactly the
situation that we described in the previous case and in fact both 10-d
configurations have the same 11-d origin, they differ only in the
choice of the $11^{th}$ coordinate.


\subsection*{Acknowledgments}


We would like to thank Ilka Agricola, Bernard de Wit and Thomas
Friedrich for useful discussions and especially Paul Saffin for
spotting a number of typos in the earlier version of this paper.
K.B. would like to thank the Theory group of the University of Milan,
where part of this work has been carried out.  The work of K.B. is
supported by a Heisenberg grant of the DFG and the work of C.J. by a
Graduiertenkolleg grant of the DFG (The Standard Model of Particle
Physics - structure, precision tests and extensions).




\providecommand{\href}[2]{#2}\begingroup\raggedright\endgroup

\end{document}